# Direct evidence of a continuous transition between waves and particles

Christian Kisielowski[1]‡, Petra Specht[2], Stig Helveg[3], Fu-Rong Chen[4], Bert Freitag[5], Joerg Jinschek[6], Dirk Van Dyck[7]

[1] The Molecular Foundry, Lawrence Berkeley National Laboratory; One Cyclotron Rd., Berkeley, CA 94720, USA
[2] Department of Materials Science and Engineering, University of California Berkeley; Berkeley, CA 94720, USA
[3] Center for Visualizing Catalytic Processes (VISION), Department of Physics, Technical University of Denmark; Kgs. Lyngby, Denmark
[4] Department of Materials Science and Engineering, City University of Hong Kong; Kowloon Tong, Hong Kong, SAR.
[5] Thermo Fisher Scientific; Achtseweg Noord 5, 5651 GG Eindhoven, The Netherlands
[6] National Centre for Nano Fabrication and Characterization (DTU Nanolab), Technical University of Denmark, Kgs. Lyngby, Denmark
[7] Electron Microscopy for Materials Science (EMAT); University of Antwerp, 2020 Antwerp, Belgium

**Abstract**

The correlation between particle and wave descriptions of electron-matter interactions is analyzed by measuring the delocalization of an evanescent field using electron microscopy. Its spatial extension coincides with the energy-dependent, self-coherence length of propagating wave packets that obey the time-dependent Schrödinger equation and undergo a Goos-Hänchen shift. In the Heisenberg limit they are created by self-interferences during coherent-inelastic Coulomb interactions with a decoherence phase $\Delta\varphi = 0.5$ rad and shrink to particle-like dimensions for energy losses of more than 1000 eV.

PACS: 03., 03.75.-b, 34.80.Pa, 07.78.+s,

The Copenhagen convention of quantum mechanics [1] describes a boundary between the quantum world, where wave functions are indefinite in space and time, and the classical world of particles that we perceive as definite. Transmission electron microscopy (TEM) is the ideal method to explore this boundary as it can reveal the underlying wave/particle dualism [2]. Nowadays, this duality can be detected in every interference experiment at the ultimate detection limit, where isolated electron scattering events are directly observed in high-performance electron microscopes [3]. They prove that electron wave functions collapse and become particle-like when detected in the classical world. However, particle or wave models describing Coulomb interactions of electrons traversing matter often seem disconnected, being expressed either by an exchange of virtual photons in a pure particle picture (Fig. 1a) or by Huygen's Principle for coherent-elastic interferences in a pure wave picture (Fig. 1b, $\Delta E=0$). For decades, the latter includes the multi-slice approach [4] that reliably solves Schrödinger's equation to simulate the formation of atomic resolution images. Herein, coherence is mandatory for an image formation by wave interferences and described by a temporal (longitudinal) coherence length and a spatial (transverse) coherence length for any ensemble of scattered electrons that is used for imaging. $\lambda$ is the de Broglie wavelength with uncertainty $\Delta\lambda$ that is commonly attributed to the energy spread of the electron source and $\theta$ is the angular spread of the probing electron beam [5]. In current TEM image simulations, a partial coherence of the electron ensemble is included in damping functions to the contrast transfer function of the microscopes' objective lens ignoring an impact on isolated electron scattering events. Moreover, in textbook considerations, the electron scattering is perceived as either coherent-elastic or incoherent-inelastic [5]. However, there is growing evidence that Coulomb interactions are always inelastic [6, 7] and definitions of coherence or decoherence remain ambiguous. Consequently, the description of electron scattering must be considered carefully [8] with the aim to explain diverging interpretations [9-12]. Certainly, the entanglement of electron source and detector is also critically relevant to understand dynamical or quantum mechanical investigations [13-15] with a spatiotemporal resolution that approaches 1 Ångstrøm at 1 picosecond or less [16].

To investigate the nature of Coulomb interactions the TEAM I microscope is employed, which was developed by the Department of Energy to promote deep sub-Ångstrom resolution by advancing aberration correction and camera performance [17,18]. Briefly, the microscope is operated in a Nelsonian illumination mode that is suitably generated by a monochromator/gun assembly, with chromatic (Cc) and spherical (Cs) aberration correction, and a direct electron detector (K2 camera). It produces a pencil-like, highly coherent electron beam irradiating a $\sim 10^7$ Å$^2$ large field of view, which is captured on the direct electron detector.

For the current experiments, a K3 camera was mounted behind a Gatan Image Filter (GIF) to acquire energy filtered transmission electron micrographs (EFTEM). Electrons accelerated by 300 kV have a de Broglie wavelength of 2 pm and travel at 78 percent of the speed of light. Their wavelength increases by $\Delta\lambda$ if energy losses $\Delta E$ occur during their interactions with solids. The instrumental energy resolution is set by the energy spread of the electron source and the energy selecting slit width of the GIF. Two microscope settings are calibrated for imaging with electron beams of energy spread $E_{FWHM}$= 0.6 eV and $E_{FWHM}$= 0.1 eV at full width half maximum together with slit widths of 1 or 2 eV and 0.3 eV, respectively. The conditions differ in terms of irradiated areas and beam coherence, where either the $C_2$ condenser aperture determines the field of view or the aforementioned Nelsonian illumination scheme. Tuning the energy filter and gain correction of the camera are performed by standard procedures to minimize intensity variations across the field of view that become noticeable for energy losses below ~ 1 eV. The presence of a Cc corrector ensures that all coherent-inelastically scattered electrons with energy losses below ~ 600 eV are focused in the same imaging plane [19]. We pursue amplitude imaging with defocus $\Delta f = 0$ nm, which is established by minimizing the Fresnel phase contrast fringes at sample/vacuum interfaces (Fig. S1, S2a) [20]. For each energy setting we record the incident electron distribution I by capturing single electron scattering events in an empty image frame and the scattered intensity $I_s$ by shifting the sample into the field of view without changing any other acquisition parameters. The sample was located such that the image covered in the broad-beam TEM mode both the sample and vacuum regions in projections. The resulting difference images ($I_s$-I) are used to quantitatively probe spatial intensity distributions. All image processing, Pendellösung, and multi-slice calculations are performed with the Tempas software package [21].

The free standing GaN samples of our experiments are of wurzite structure and produced by a focused ion beam (FIB) milling process that is followed by cleaning with low energetic (500 V) argon ions. The process yields plane parallel samples together with a wedge perpendicular to the c-planes of the sample and an abrupt m-plane sample/vacuum interface as shown in Fig. 2a. A sample thickness of 290 nm is determined by calculating that Pendellösung oscillations are of 48 nm periodicity, which form the six visible contrast oscillations along the c-planes. Further, our sample cleaning process removes all residual amorphous surface contaminations to produce atomically stepped surfaces (inset Fig. 2a). The two EFTEM amplitude images of the non-polar GaN m-plane in Fig. 2b highlight the evanescent electron intensity that extends from the sample into the vacuum across a distance $x_i$, which is of interest here.

We report that $x_i$ systematically shrinks with increasing energy loss energy loss as shown in the Fig. 2b and in Fig. S2b, S3a for other investigated energy losses [20]. Current multi-slice simulations of amplitude images from aberration-corrected electron microscopes do not reproduce this extension of image intensity into the vacuum (Fig. 1c). Further, single pixel line profiles across the GaN/vacuum interface are entirely dominated by telegraph noise because quantized scattering events are detected with an outstanding signal-to-noise ratio (Fig. 3a). For noise reduction, image series up to 100 images are recorded, aligned, and averaged. To generate a final profile as shown in Fig. 3b, we additionally averaged 500 - 1000 pixels parallel to the abrupt interface. Thereby, it is uncovered that the intensity extension into the vacuum is strictly exponential $\sim e^{-x/x_i}$ if single electron scattering events are detected. In addition, the observation is independent of the accumulated electron dose, which excludes an impact of beam damage or the number of the detected electrons on the decay characteristics. It only exhibits the expected limits to spatial resolution caused by a finite number of electron counts. For the utilized electron ensemble of Fig. 3b, a resolution limit for $x_i$ of 1.9 Å is estimated from the error of fitting an exponential decay. Surprisingly, all measured energy dependent values $x_i(\Delta E)$ for energy losses between 40 eV and 0.9 eV rigorously follow the simple law $x_i(\Delta E) = 1/a\Delta E$, which is independent of the energy spread of the electron source (Fig. 4a).

Two systematic but previously unknown dependencies of the data sets are striking: First, while exponential intensity decays in surface proximity such as shown in Figure 3b are generally considered e.g. Ref. [22] there is an unambiguous $1/\Delta E$ dependence of the decay constants $x_i(\Delta E)$ on the reciprocal energy (Fig. 4a) for single electron self-interferences. Second, the data $x_i(\Delta E)$ follow closely the $1/\Delta E$ dependence of Heisenberg's Uncertainty Principle when simply added to Fig. 4b, which was derived earlier [3].

Both aspects can be rationalized by noting that the superposition principle is one of the strongest principles in quantum mechanics with a remarkable link to self-interferences and decoherence [3]. In this picture, the formation of wave packets is explained by the coexistence of wave functions with wavelength differences $\Delta\lambda$ that self-interfere during Coulomb interactions with an energy exchange $\Delta E$ (Fig. 1b). In this case, a self-coherence length $l_s$ is expressed from a phase (path) difference $\Delta\varphi$ of two partial waves by:

(1)
$$l_S = \frac{\Delta\varphi}{\frac{2\pi}{\lambda} - \frac{2\pi}{\lambda + \Delta\lambda}}$$

For $\Delta\varphi = 1$ rad, the description reproduces the established longitudinal coherence criterion $l_l = \lambda^2/\Delta\lambda$ that applies to photon or electron ensembles. The significance of choosing a phase shift $\Delta\varphi = 0.5$ rad for self-interferences becomes evident when the calculated self-coherence length is divided by the speed of light c to give a self-coherence time $t_s = l_s/c$ as reproduced in Figure 4b. Since the slope of the straight line $\Delta t_s \Delta E$ is given by $h/4\pi = \hbar/2$, where h is the Planck constant, the depicted linear relation describes Heisenberg's Uncertainty Principle $\Delta t \Delta E \geq \hbar/2$ by self-interferences. Using the decoherence phase $\Delta\varphi = 0.5$ rad, it marks the boundary between classical mechanics (white background) and quantum mechanics (gray background). In this picture, the self-coherence length of scattered electrons shrinks rapidly with increasing energy losses associated with phonon, plasmon, or core excitations in the sample. Because the wave packets continuously narrow as the strength of the Coulomb interaction increases, they become increasingly particle-like and can be considered collapsed in the classical world when $l_s$ reaches atomic dimensions.

From an optical perspective, the shallow incident angles at sample/vacuum interfaces implies that total internal reflection of waves occurs and, in turn, creates a transversal evanescent field within a coherently illuminated area as explained by the longitudinal Goos-Hänchen Shift (GHS) D (Fig.1c) [23]. A quantum mechanical equivalent of optical pulses is explicitly debated by Berman [24]. Specifically, it is established that the time-dependent Schrödinger equation with wave functions $\psi(r, t) = \psi(r) \varphi(t)$ has time dependent solutions for pulses of both electromagnetic and matter waves that create transversal evanescent fields of decay length $l_e$, which we estimate by [24]:

(2)
$$l_e(\Delta E) \cong |D| = \left|\frac{\lambda}{\pi} \frac{sin(\Phi)}{\sqrt{sin(\Phi)^2 - n^2}}\right| = \frac{\hbar c}{\Delta E}$$

Here, $\lambda$ is either the de Broglie or an electromagnetic wavelength. Estimating $\lambda$ by the uncertainty $\Delta\lambda = hc/\Delta E$ assumes a virtual photon exchange between electron and sample during the Coulomb interaction, which relates the quantum mechanical and classical worlds. Using $n^2 = 5$ for GaN and $\Phi = \pi/2$ in a wave picture the expression yields the close agreement $x_i \approx l_s \leq l_e$ in

Fig. 4b. The GHS enables our measurements at sample/vacuum interfaces because it provides a stationary, transversal solution for the time-dependence of Schrödinger's equation

$$\overline{\varphi(t)} \sim e^{-\frac{i \Delta E t}{\hbar}}$$

(equation S7) [20] with a strict intensity decay $e^{-x/x_i(\Delta E)}$ suggesting

(3)
$$x_i = \frac{\hbar x}{t \Delta E} = \frac{\hbar v}{\Delta E}$$

where v is the propagation velocity of a wave packet. The expression is in close agreement with the estimated magnitude of the GHS.

However, equation 3 explains the dependence of the penetration distance on the reciprocal energy in Fig.4b by the first derivative of Schrödinger's equation with respect to time. Wave packets propagating at the speed of light

$$x_{ic} = l_e = \frac{\hbar c}{\Delta E}$$

include the fundamental constant $\hbar c$ =197 eVnm, which is approached by ~ 54% in our experiments with GaN. A reduced propagation velocity in the medium also explains the small but systematic deviation from Heisenberg's relation in Fig. 4b. This intriguing similarity of wavelengths and the decoherence of self-interfering wave functions attributes the reciprocal dependence of $x_i$ on energy to the propagation of wave packets in the time domain and the constant magnitude of the GHS at large scattering angles to the self-coherence length. Quantum mechanical tunneling behavior is inconsistent with our observations because of its reciprocal square root dependence on energy (Fig 4b). This originates from the second derivative of Schrödinger's equation (equation S5) [20]. In addition, the pre-factor ψ(r) characterizes the electron energy loss spectrum (Figure S3b) [20], which is nowadays accessible down to energy losses in the phonon region [25] and often exploited in aloof geometry [26, 27]. Such experiments are fully compatible with our results as is the recent assessment that phonon excitations dominate if ultra-low dose imaging conditions apply (Fig. 4b). Many observations in TEM experiments can successfully be described by image simulations based on the above-mentioned multi-slice approximation [4]. However, for a correct description of observations with high spatiotemporal resolution, we must include pulse-like, inelastic beam-sample interactions, the impact of scattering on self-coherence, and potentially an entanglement of electron source and detector.

Concluding, energy dependent electron-matter interactions generate wave packets by self-interferences that continuously transform the quantum mechanical wave picture into a particle picture. The transition is observed in a finite detection window because the wave packets collapse below the spatial resolution limit of the electron microscope (0.5 Å) for high energy losses ΔE **>** 1000 eV while a finite energy resolution of 0.3 eV ($E_{slit}$) or residual aberrations of the GIF masks the GHS at low energies ΔE ≲ 0.9 eV. In this model, coherent-inelastic interactions inevitably generate pulse-like wave packets at any energy loss , which must be considered to explain electron scattering beyond static models because they stimulate structural dynamics across a vast time scale [13, 14, 16]. Further, it is established that photonic excitations of structural dynamics can alter the pristine atom configurations of solids, often exceeding equivalent excitations caused by an electron beam [28]. Therefore, a "divide and conquer" approach for electrons exploits intrinsic advantages by working with ultra-low beam currents (< 50 femto Amperes) delivered into a well-defined area [3, 7] and by structuring the electron beam in the time domain [16]. It provides an opportunity to bring together best practices developed for electron microscopy in biological, chemical, and material sciences, by preferably using phonon excitations to form images, which finds a spectroscopic counterpart in aloof experiments that exploit the spatial extension of the evanescent field. Importantly, it is feasible to expand this view beyond Coulomb forces and to account for any virtual particle exchange between coexisting wave functions that characterize other interactions.

**Acknowledgments:**

Electron microscopy is supported by the Molecular Foundry, which is supported by the Office of Science, the Office of Basic Energy Sciences, the U.S. Department of Energy under Contract No. DE-AC02-05CH11231. Investigations of GaN are supported by DTRA contract #HDTRA1-17-1-0032. The Center for Visualizing Catalytic Processes is sponsored by the Danish National Research Foundation (DNRF146). FRC is supported by the project HZQB-KCZYB-2020031.

[‡]CFKisielowski@lbl.gov

# Figures

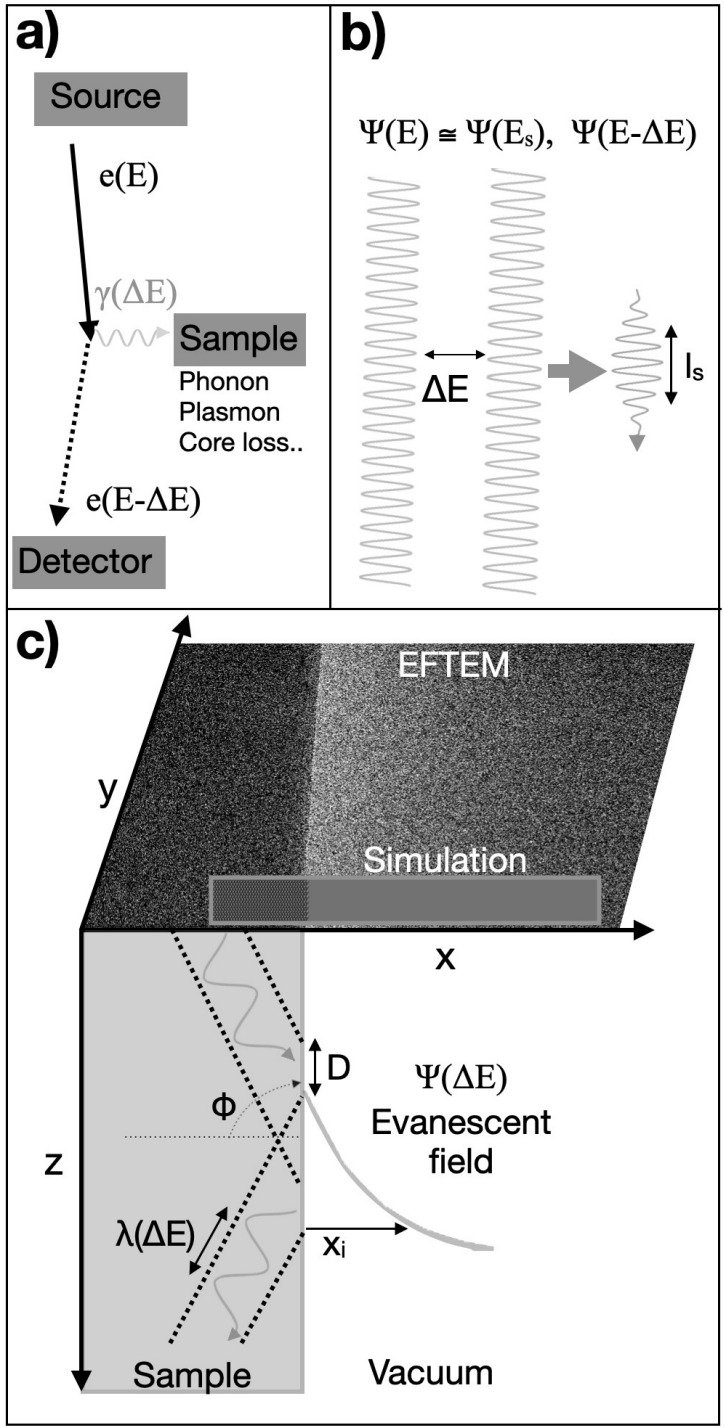

**Fig. 1. Schematics of electron scattering in particle and wave pictures and their relation to the performed measurements.**

a) Particle picture. e(E) = emitted electron of energy E = 300 keV (travels at 78 % of light speed); γ(ΔE) = virtual photon exchange during Coulomb interaction, which excites particles in the solid; e(E-ΔE) = detected electron.

b) Wave picture: The emitted wave function Ψ(E) and the scattered wave function $\Psi(E_s)$ coexist and are of similar energy/wavelength. For elastic interactions ΔE = 0 all wave functions remain delocalized in space and time. For inelastic interactions a small energy portion $\Delta E = E - E_s \neq 0$ (in the range of eV) is exchanged during scattering. Thereby, a wave package Ψ(E-ΔE) can be formed by self-interference with a finite self-coherence length $l_s$ (equation 1) that propagates in z-direction.

c) Measurement: Experimentally we measure wave packets Ψ(ΔE) that propagate in beam direction. For incidence angles Φ close to $90^0$ they experience a total internal reflection at sample/vacuum interfaces together with a Goos-Hänchen shift D that establish a transversal, exponential decay of an evanescent field (equation 2). The energy dependent penetration distance $x_i$ (delocalization) of the probability distribution I=ΨΨ* is measured in energy filtered TEM images (EFTEM). A multi-slice simulation is inserted.

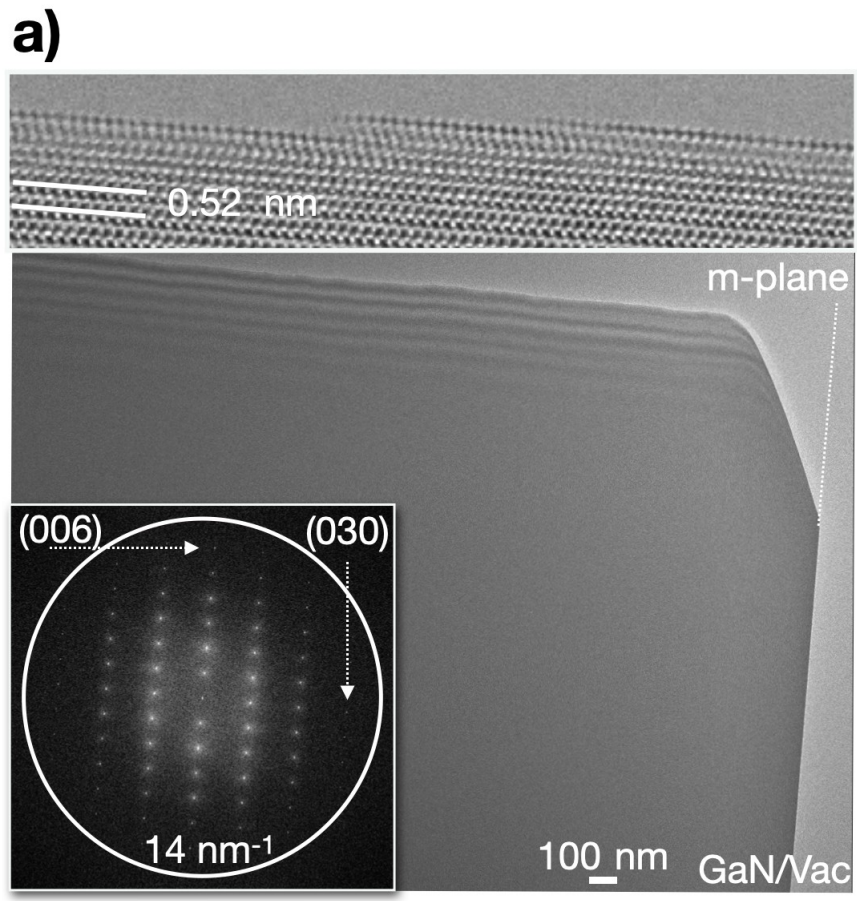

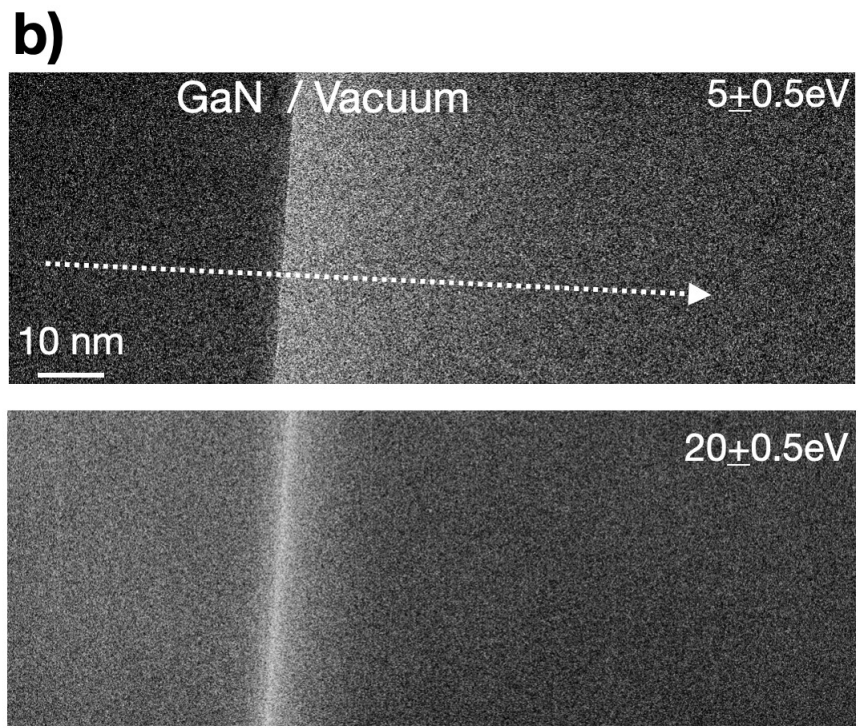

**Fig. 2. Characterization of the plane-parallel GaN sample and energy filtered transition electron micrographs (EFTEM images).**

a) A low magnification image of the investigated GaN sample is shown in a-plane orientation together with a high-resolution image of the local wedge at the top of the sample (c-plane) and its Fourier transform (inset). Surface contamination layers are absent. In this work, the bright contrast surrounding the sample on a scale of 10 nm is investigated at the abrupt, non-polar m-plane/vacuum interface.

b) Recorded EFTEM amplitude images with two different energy losses as indicated. The dotted line marks the direction for the extraction of the intensity profile of Figure 3a.

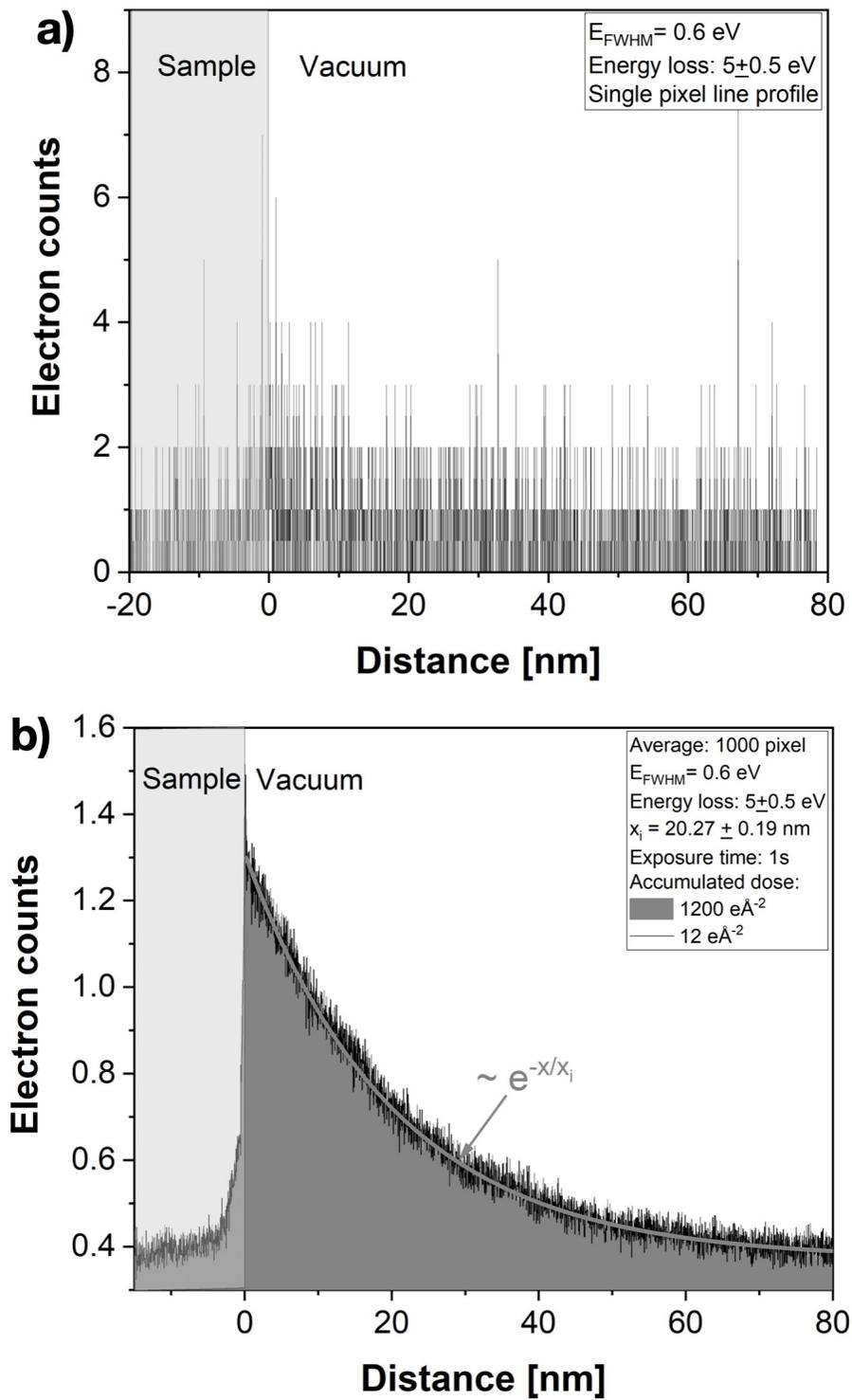

**Fig. 3. Single electron detection and analysis.**

a) Single electron scattering events are detected with an outstanding S/N ratio. The resulting telegraph noise dominates the recorded EFTEM images (Fig. 2b) and is quantified by the line profile of single pixel width.

b) Averaging along the interface and the recording of image series are used for noise reduction. The intensity decays strictly exponential with a characteristic distance $x_i$ that we determine for a range of energy losses to obtain the data sets shown in Fig. 4a. Decay characteristics are independent of those accumulation, which is altered by a factor 100 in this example. Averaging reduces the experimental errors of $x_i$ to below 1% or 1.9 Å as indicated, which is an estimate for the obtained resolution.

**Fig. 4. Energy dependence of the penetration depth and its relation to self-coherence.**

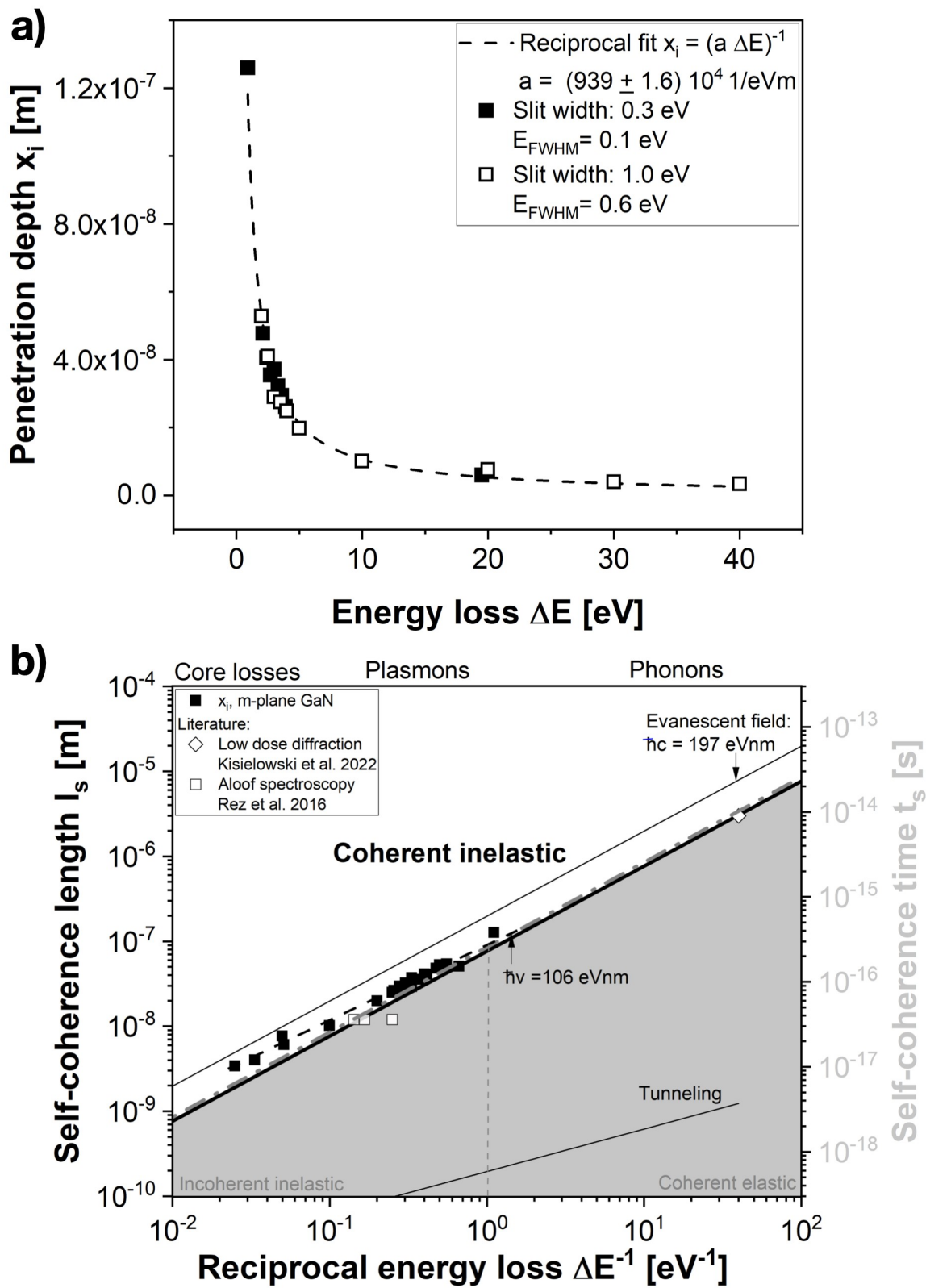

a) A strict dependence $x_i(\Delta E) = 1/a\Delta E$ is uncovered. Two different illumination conditions are used as indicated. The fitting parameter a relates to the propagation velocity v of wave packets by 106 eVnm = $\hbar v$ = 1/a.

b) Spatial (bold solid black) and temporal (bold dash-dot gray) self-coherence versus reciprocal energy loss in coherent-inelastic electron scattering [3]. Heisenberg's Uncertainty Principle is matched by choosing a decoherence phase of $\Delta \varphi = 0.5$ rad. The values $x_i(\Delta E)$ of Fig. 4b (black squares) are simply added to the graph together with results from Ref. [7, 24] (open squares). They are compared to the self-coherence lengths $l_s$ (equation 1), tunneling distances $l_t$ (SI, equation S5), and to the extension of evanescent fields $l_e$ created by wave-packets traveling at the speed of light (equation 2, $\hbar c$ = 197 eVnm) and at 54% of light speed (equation 3, $\hbar v$ = 106 eVnm). Traditionally, an abrupt coherence change around 1 eV is postulated to model either coherent-elastic or incoherent-inelastic scattering events.

# Supplementary Information

## Direct evidence of a continuous transition between waves and particles

Christian Kisielowski, Petra Specht, Stig Helveg, Fu-Rong Chen, Bert Freitag, Joerg Jinschek, Dirk Van Dyck

Correspondence to: CFKisielowski@lbl.gov

**This file includes:**
　　Supplementary equations
　　Supplementary Figs. S1 to S3

**Supplementary Figures:**

**Fig. S1**

Energy filtered transition electron microscopy (EFTEM). The incident electron distribution is not subtracted.

a) Top: An as recorded bright field image of the GaN/vacuum interface with a multi-slice simulation of an amplitude image inserted. Bottom: EFTEM image of the same area by filtering the zero-beam only at $(0 \pm 0.5)$ eV.

b) Averaged intensity profiles perpendicular to the interface. Fresnel fringes occur at the interface if the defocus $\Delta f$ deviates from zero and are suppressed in amplitude images by choosing $\Delta f = 0$ as shown by the multi-slice simulation. In energy filtered images with losses below ~ 1 eV, the electron counts increase exponentially with the distance from the surface. This effect is unconsidered in this paper and we speculate that it relates to an entanglement of source and detector.

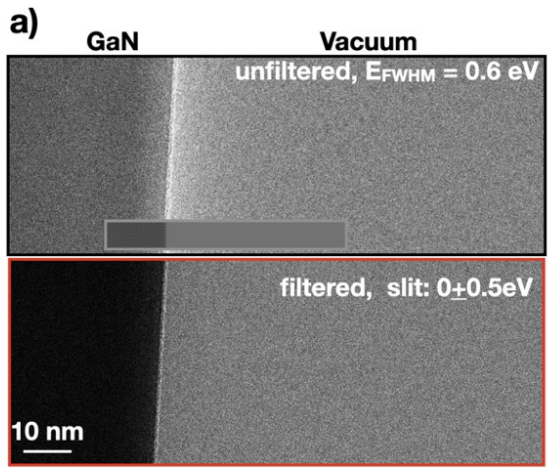
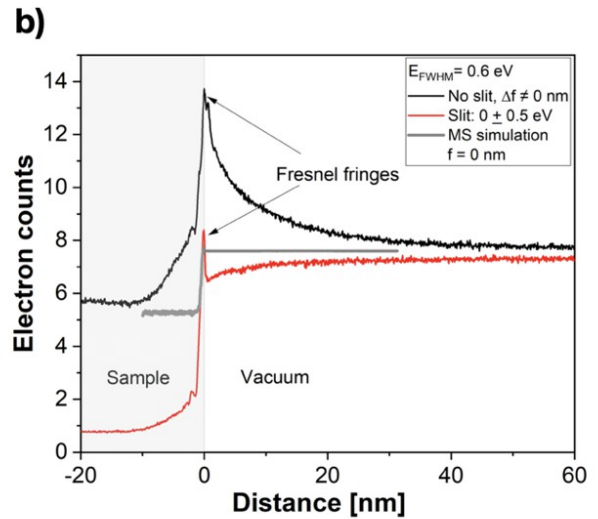

**Fig. S2**

Image intensity variations in the energy loss region. The incident electron distribution is subtracted.

a) A transition between an exponential intensity growth and an exponential intensity decay occurs in the amplitude images of the vacuum at low energy losses ≤ 2 eV for a beam spread of $E_{FWHM}$= 0.6 eV and a slit width slit width of 2 eV. Fresnel fringes are absent because $\Delta f=0$.

b) Normalized intensity decay for selected energy losses ≥ 0.9 eV that are analyzed in this paper. Recording with an electron beam of $E_{FWHM}$= 0.1 eV and a narrow slit (0.3 eV) using the Nelsonian illumination mode. $I_o$= intensity at the interface. The decay is strictly exponential across the investigated energy range.

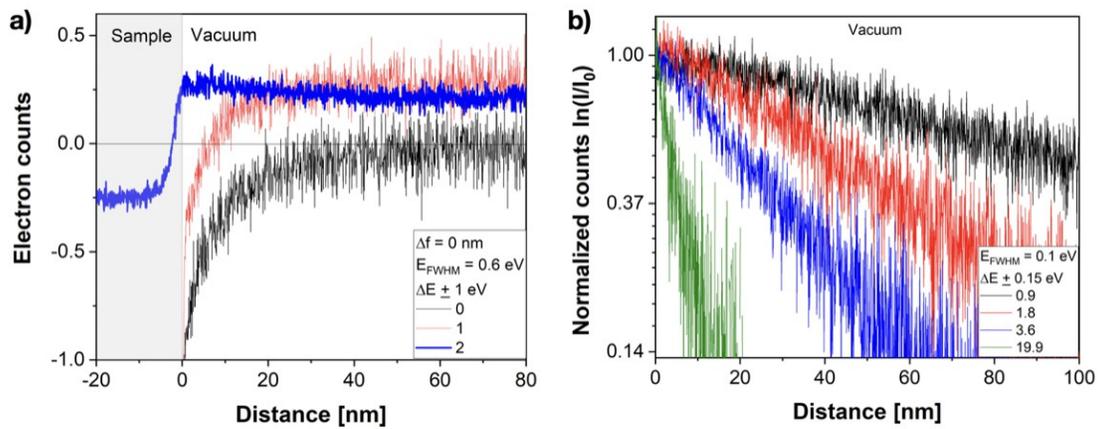

**Fig. S3**

Image intensity decay in the energy loss region and the relation to electron energy loss spectroscopy (EELS). The incident electron distribution is subtracted.
a) Normalized exponential decay at selected energy losses ≥ 2.5 eV recorded with an electron beam of $E_{FWHM}$= 0.6 eV and a slit width of 1 eV using the traditional illumination mode.
b) The electron count at the interface mimics the electron energy loss spectrum of GaN at a poor resolution except for energies close to the beam width. An EELS spectrum of GaN is inserted. Plasmon losses in GaN peak at the characteristic energy value of 19.4 eV.

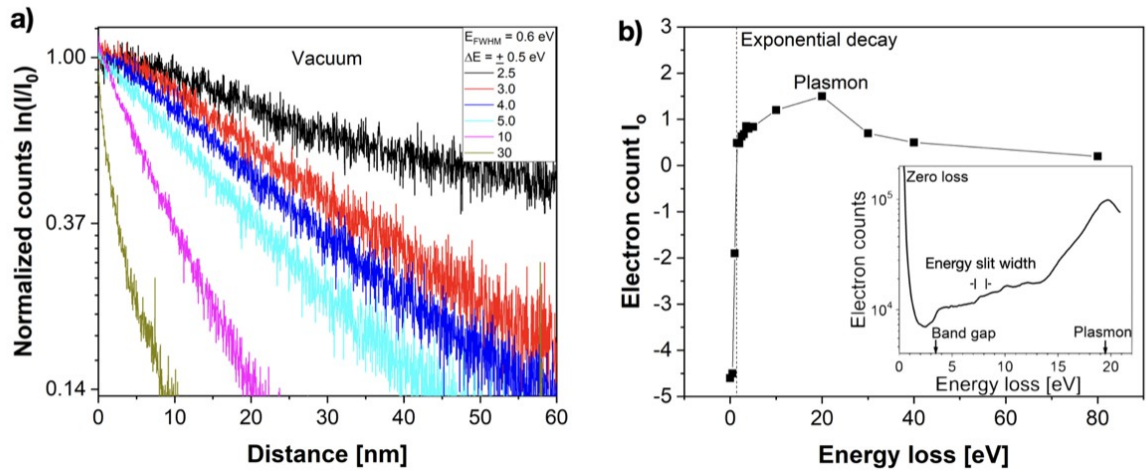

**Supplementary equations**

The time dependent Schrödinger equation is discussed for electron microscopical applications.[5] Here we briefly highlight features at are relevant to this publication. The equation is given by:

$$(S1) \quad -\frac{\hbar^2}{2m}\nabla^2\psi(r,t) + V(r)\psi(r,t) = i\hbar\frac{d\psi(r,t)}{dt}$$

It is common practice to obtain solutions by a separation of variables ψ(r,t) = ψ(r) φ(t), which yields:

$$(S2) \quad -\frac{\hbar^2}{2m}\frac{1}{\psi(r)}\nabla^2\psi(r) + V(r) = E = i\hbar\frac{1}{\varphi(t)}\frac{d\varphi(t)}{dt}$$

where the constant E is an energy value. Thus, the time independent part of the solution obeys the differential equation:

$$(S3) \quad \nabla^2\psi(r) = -\frac{2m}{\hbar^2}\big(E - V(r)\big)\psi(r) = -k^2\psi(r)$$

For bound solutions such as a particle in a one dimensional box the wave function penetrates the potential wall and decays exponentially as:

$$(S4) \quad \psi(x) = Be^{-kx}$$

$\frac{1}{k} = l_t$ is the tunneling distance.

$$(S5) \quad l_t(\Delta E) = \frac{\hbar}{\sqrt{2m\,\Delta E}}$$

It is seen that a square root dependence on the inverse energy originates from the second derivative (curvature) of the time independent Schrödinger equation. It is inconsistent with our observations in Fig.4a.

On the other hand, the time dependent part of equation S2 obeys the differential equation:

$$(S6) \quad \frac{d}{dt}\varphi(t) = \frac{E}{i\hbar}\varphi(t)$$

which solves as:

$$(S7) \qquad \varphi(t) = \varphi(0)e^{\frac{-iEt}{\hbar}}$$

Here, $t_i(E) = \dfrac{\hbar}{E}$ is an energy dependent time constant that can be transformed to a length scale by multiplication with the speed of light. Thereby, it is seen that a linear dependence of the decay constant on the inverse energy originates from the first-time derivative (slope) of Schrödinger's equation. It is consistent with our observations in Fig.4b. In addition, it is reported that the Goos-Hänchen shift causes the formation of a transversal, static evanescent field by propagating quantum mechanical wave packages and optical pulses.[23] Its classical equivalent[22] is well-known from information transfer in fiber optics or total reflection microscopy.